\documentclass[aps,prl,reprint,superscriptaddress,nobibnotes]{revtex4-2}
\usepackage{blindtext}
\usepackage{lineno}
\usepackage{nicefrac}
\usepackage{amsmath}
\usepackage{graphicx}
\usepackage{epstopdf}
\usepackage{float}
\usepackage{siunitx}
\usepackage{tabularx}
\usepackage{txfonts}
\usepackage[utf8]{inputenc}
\usepackage[T1]{fontenc}
\usepackage{soul}

\usepackage[dvipsnames]{xcolor}

\colorlet{mylinkcolor}{RoyalPurple}
\colorlet{mycitecolor}{RoyalPurple}
\colorlet{myurlcolor}{RoyalPurple}
\usepackage{hyperref}
\hypersetup{
	linkcolor  = mylinkcolor,
	citecolor  = mycitecolor,
	urlcolor   = myurlcolor,
	colorlinks = true,
	breaklinks = true
}

\DeclareSIUnit{\rad}{rad}

\newcommand{\subfig}[1]{(#1)}

\global\long\def\infint#1{\intop\:\mathrm{d}#1}%
\global\long\def\et{\psi}%
\global\long\def\d{\mathrm{d}}%
\global\long\def\frft{\widehat{\mathrm{FrFT}}^{(\varphi)}}

\begin{document}
\title{Experimental implementation of the optical fractional Fourier transform in the time-frequency domain}
\author{Bartosz Niewelt}
\thanks{Equal contributions}
\affiliation{Centre for Quantum Optical Technologies, Centre of New Technologies, University of Warsaw, Banacha 2c, 02-097 Warsaw, Poland}
\affiliation{Faculty of Physics, University of Warsaw, Pasteura 5, 02-093 Warsaw, Poland}
\author{Marcin Jastrzębski}
\thanks{Equal contributions}
\affiliation{Centre for Quantum Optical Technologies, Centre of New Technologies, University of Warsaw, Banacha 2c, 02-097 Warsaw, Poland}
\affiliation{Faculty of Physics, University of Warsaw, Pasteura 5, 02-093 Warsaw, Poland}
\author{Stanisław Kurzyna}
\thanks{Equal contributions}
\affiliation{Centre for Quantum Optical Technologies, Centre of New Technologies, University of Warsaw, Banacha 2c, 02-097 Warsaw, Poland}
\affiliation{Faculty of Physics, University of Warsaw, Pasteura 5, 02-093 Warsaw, Poland}
\author{Jan Nowosielski}
\thanks{Equal contributions}
\affiliation{Centre for Quantum Optical Technologies, Centre of New Technologies, University of Warsaw, Banacha 2c, 02-097 Warsaw, Poland}
\affiliation{Faculty of Physics, University of Warsaw, Pasteura 5, 02-093 Warsaw, Poland}
\author{Wojciech Wasilewski}
\affiliation{Centre for Quantum Optical Technologies, Centre of New Technologies, University of Warsaw, Banacha 2c, 02-097 Warsaw, Poland}
\affiliation{Faculty of Physics, University of Warsaw, Pasteura 5, 02-093 Warsaw, Poland}
\author{Mateusz Mazelanik}
\email{m.mazelanik@cent.uw.edu.pl}
\affiliation{Centre for Quantum Optical Technologies, Centre of New Technologies, University of Warsaw, Banacha 2c, 02-097 Warsaw, Poland}
\affiliation{Faculty of Physics, University of Warsaw, Pasteura 5, 02-093 Warsaw, Poland}
\author{Michał Parniak}
\email{m.parniak@cent.uw.edu.pl}
\affiliation{Centre for Quantum Optical Technologies, Centre of New Technologies, University of Warsaw, Banacha 2c, 02-097 Warsaw, Poland}
\affiliation{Niels Bohr Institute, University of Copenhagen, Blegdamsvej 17, 2100 Copenhagen}

\begin{abstract}
The fractional Fourier transform (FrFT), a fundamental operation in physics that corresponds to a rotation of phase space by any angle, is also an indispensable tool employed in digital signal processing for noise reduction. Processing of optical signals in their time-frequency degree of freedom bypasses the digitization step and presents an opportunity to enhance many protocols in quantum and classical communication, sensing and computing. In this letter, we present the experimental realization of the fractional Fourier transform in the time-frequency domain using an atomic quantum-optical memory system with processing capabilities. Our scheme performs the operation by imposing programmable interleaved spectral and temporal phases. We have verified the FrFT by analyses of chroncyclic Wigner functions measured via a shot-noise limited homodyne detector. Our results hold  prospects for achieving temporal-mode sorting, processing and super-resolved parameter estimation.

\end{abstract}

\maketitle

\paragraph{Introduction}

Harnessing many degrees of freedom of photons, such as polarization \cite{han_coherent_2005}, spatial modes, in particular with orbital angular momentum, or temporal modes \cite{brecht_photon_2015} holds continued importance in quantum protocols where researchers seek to achieve new capabilities or gain enhanced performance and capacity. The time-frequency domain is exploited to a great effect, but typically only by using parallel spectral channels, as witnessed by the wavelength-division multiplexing devices \cite{Eriksson2019,Choi:14,Peters:10,Bahrani2015}. Nevertheless, from the perspective of quantum optics, it is clear that more advanced spectro-temporal encoding and manipulations lead to enhanced performance. For example, one may implement quantum gates and beamsplitter operations between spectral or temporal modes \cite{brecht_photon_2015,Humphreys2013,Campbell2012}. This discrete-mode picture is accompanied by a continuous treatment of the time-frequency domain \cite{Fabre2022,Fabre2023,Humphreys2014}. In particular, hybrid approaches where discrete and continuous spaces are combined are gaining attention in quantum networking \cite{Darras2023}. The continuous domain, implemented via both processing and detection with resolution exceeding the characteristic coarse-graining, often allows for accessing the highest available dimensionality of the quantum system \cite{Schneeloch2014,Tasca2013,Xiao2020}. This is particularly relevant for Einstein-Podolsky-Rosen type experiments, which provide entanglement as a resource for high-dimensional quantum key distribution \cite{Zhong2015,Nunn:13}. As far as processing is concerned, one of the fundamental operations for the continuous variables is an optical Fourier transform \cite{goodman1996introduction} that allows switching between two conjugate variables. In the spatial domain, these are the position and momentum of a photon and the Fourier transform can be achieved via a single lens placed a focal length from the conjugate planes.

The celebrated space-time duality \cite{Kolner:89,Patera2018} provides the idea of implementing the same Fourier transform in the time-frequency domain via spectral and temporal quadratic phases. A combination of those (time lenses and dispersion) finds applications in temporal imaging \cite{Kuzucu:09,Hernandez:13}, photon time-of-flight spectroscopy \cite{Goda2013} and bandwidth-matching of quantum systems \cite{Karpinski2017,Foster2009}, with implementations stretching from ultrafast optics \cite{Kauffman1994} to narrowband atomic systems \cite{Mazelanik:20}. Operations beyond the Fourier transform have been proposed but scarcely implemented. They range from trivial but useful frequency-translation operators \cite{Kurzyna:22} to complex modulations that could allow largely arbitrary modal operations \cite{HAHALU}. Those more complex operations could allow for super-resolved sensing of system parameters \cite{Mazelanik2022,Ansari2021,Donohue2018} or enhanced data rate for communication in the photon-starved regime \cite{Banaszek2020}.

\begin{figure}[ht]
\centering
\includegraphics[width = 1\columnwidth]{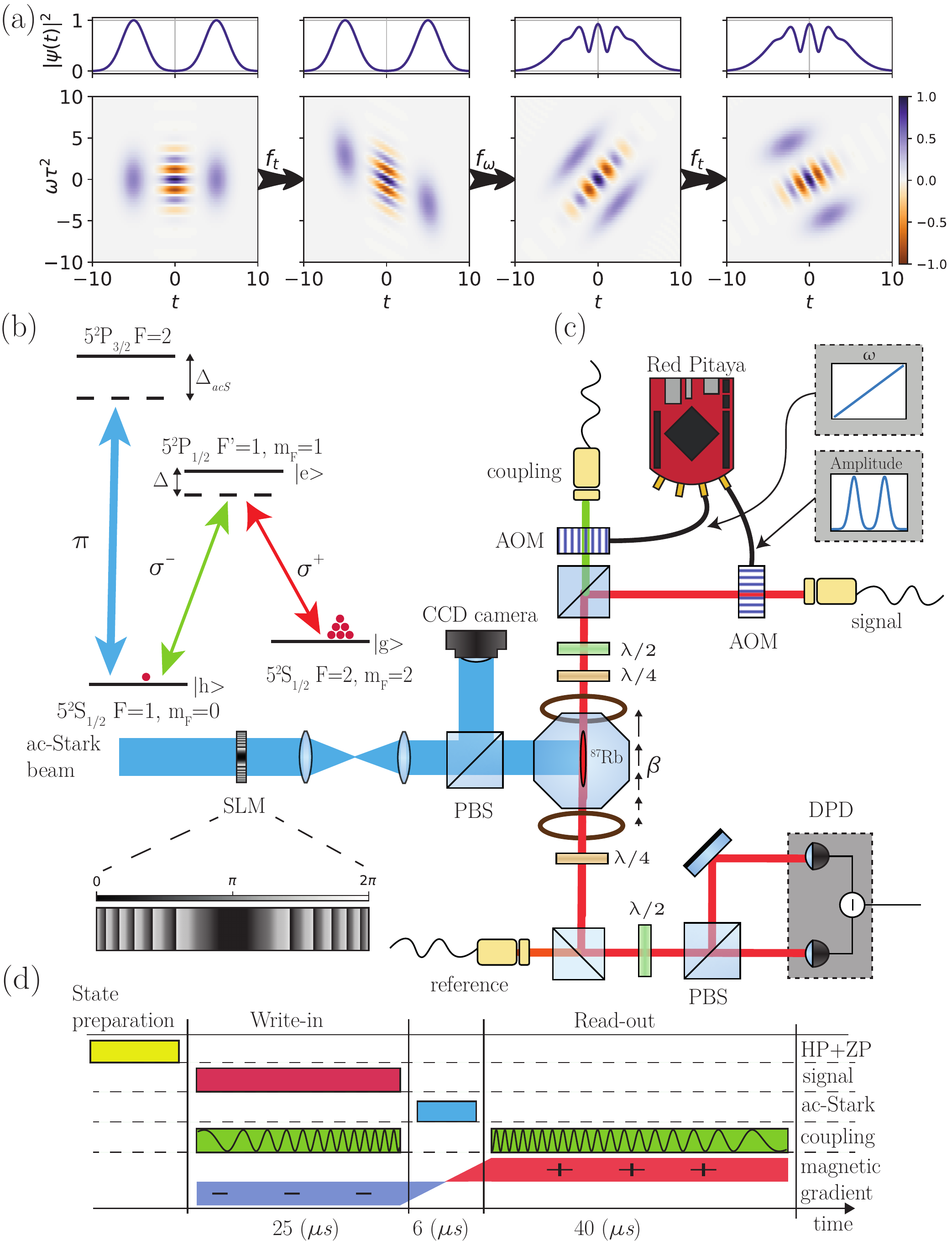}
\caption{\label{fig:idea}\subfig{a} Ideational scheme of the evolution of CWF due to temporal lenses $f_t$ and frequency lens $f_{\omega}$. \subfig{b} Relevant $^{87}\mathrm{Rb}$ energy level configuration. \subfig{c} Experimental setup used to perform FrFT, based on ultracold atoms in a magneto-optical trap. 
The frequency lens is implemented using a spatial light modulator (SLM) by imprinting an intensity pattern on ac-Stark (acS) beam. The acS beam then illuminates atoms, applying a spatial quadratic phase. The custom shot noise-limited differential photodiode (DPD) detects the beating between the signal and reference.
\subfig{d} Experimental sequence for performing FrFT of two-pulse state. HP and ZP are respectively Hyperfine pumping and Zeeman pumping.} 
\end{figure}
Here we provide an experimental implementation of the time-frequency domain optical fractional Fourier transform (FrFT). The FrFT provides a full generalization of the Fourier transform and provides significant new capabilities. Notably, in the spatial domain, the FrFT may be implemented by a different non-focal arrangement of a single lens. With this, the FrFT enables sorting of orbital angular momentum modes \cite{Zhou2017}. Optical mode sorting is of particular interest in the time-frequency domain, as it may provide temporal and spectral superresolution, enhanced optical communication with noise rejection, and non-standard coding. The more typical application of the FrFT in the engineering context is noise filtration. The FrFT is particularly well-suited to filter out structured noise, for instance with chirp \cite{Dorsch1994}, which goes beyond the typical capabilities of the conventional Fourier transform \cite{Yasuda:06}. A series of FrFTs and bandpass filters may filter out noise with complex structures, in particular using adaptive techniques \cite{Prajna2020, Krasnal}. This is typically implemented numerically, and the prospect of such noise-filtering protocols at the level of optical signals could provide significant gains for noisy communication \cite{Raymer:20}. The FrFT has also been proposed as a basis of chirp-based encoding protocols that gain rising interest in the engineering community \cite{Ouyang2016}. In physics, the FrFT in the phase-space picture arises naturally as an evolution of a quantum harmonic oscillator. 

We implement the FrFT for narrowband photons using a programmable spectral phase achieved thanks to an atomic optical quantum memory, which combines ac-Stark modulation of spin waves \cite{Mazelanik2019,Hosseini2012} and gradient-echo storage and retrieval protocol \cite{Hosseini2009, Cho:16} that inherently allows for spectral manipulation of stored signal \cite{Buchler:10,Sparkes2012}. We verify the performance of the FrFT via an illustrative example of a rotation of a coherent dual pulse, time-frequency cat-like state, and further quantify the capabilities of the device by processing Hermite-Gaussian temporal modes, which are eigenfunctions of the FrFT \cite{NAMIAS1980}. We use homodyne detection and analyze the measured chronocyclic Wigner functions (CWF) to capture the subtle characteristics of the system.

\begin{figure}[ht]
\centering
\includegraphics[width = 1\columnwidth]{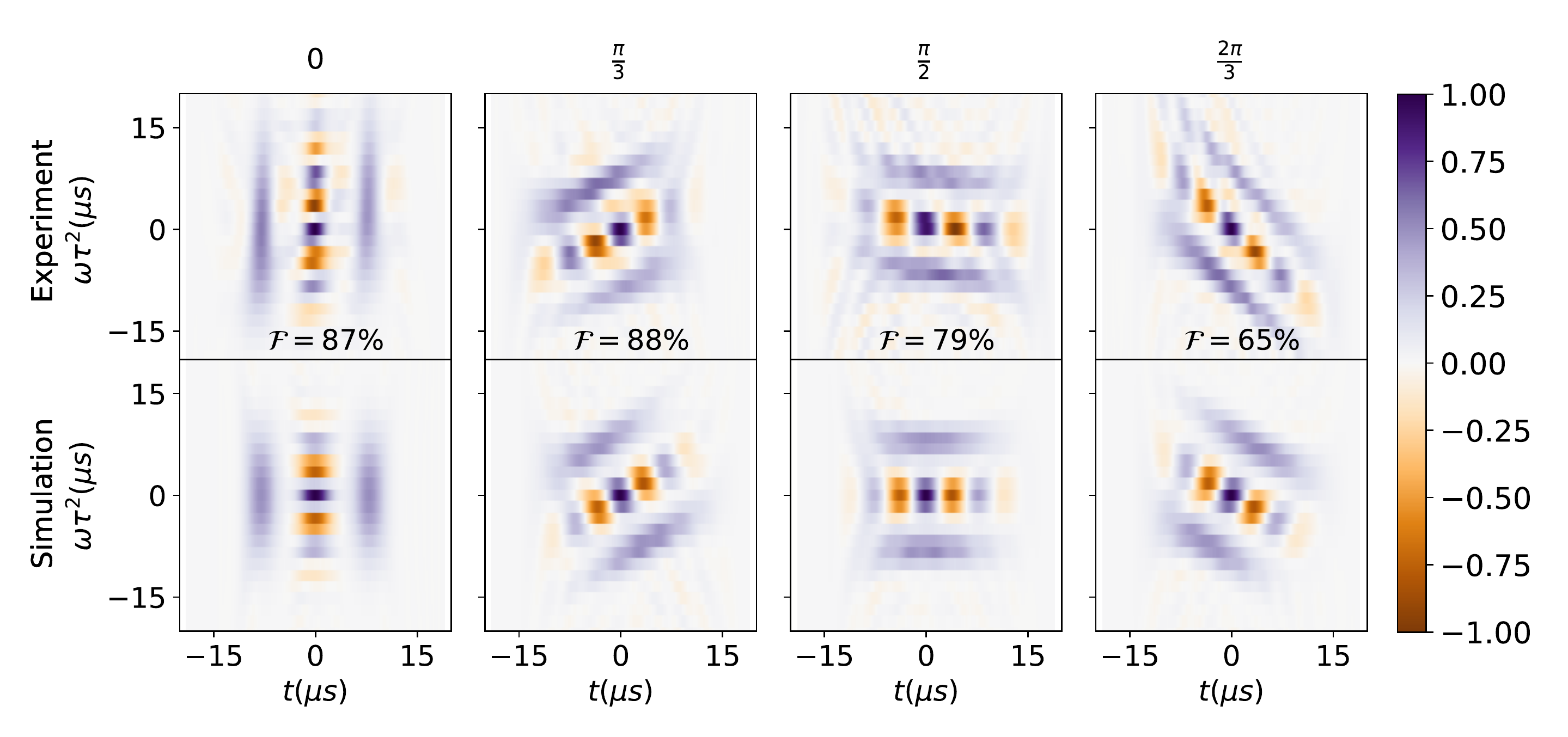}
\caption{\label{fig:rotated_data}
CWFs of two-pulse state rotated by angles $\varphi\in\{0,\frac{\pi}{3},\frac{\pi}{2},\frac{2\pi}{3}\}$. The upper row presents the experimental data. The lower row presents the results obtained from numerical simulations, taking into account experimental limitations.}
\end{figure}
\paragraph{Theory}
Let us introduce the fractional Fourier transform by the analogy to the evolution of a quantum harmonic oscillator with zero energy ground state \cite{Weimann2016}. With the corresponding Hamiltonian: $\hat{H}=\nicefrac{\hat{p}^2}{2}+\nicefrac{\hat{q}^2}{2}-\nicefrac{1}{2}$, with $\hat{p}=-i\frac{\d}{\d q}$, we define the FrFT as the evolution operator:
\begin{equation}
\frft=\exp\left(i\varphi\hat{H}\right),
\end{equation}
that evolves the initial state $\psi(q)$. The parameter $\varphi$ is then called the angle of the FrFT, and for $\varphi=\frac{\pi}{2}$ we have the Fourier transform. 
Here we deal with the spectro-temporal domain, and we take $q=\nicefrac{t}{\tau}$ and $p=\tau\omega$ as unitless quantities corresponding to the time and frequency of the optical pulses, where $\tau$ is time unit. Hence we have chronocyclic space in which we can describe the distributions using CWF, which for the electric field with complex amplitude $\et(t/\tau)$ is defined as:
\begin{equation}\label{eq:W}
W(\omega \tau,t/\tau) = \frac{1}{2\pi}\infint \xi \et\left(\frac{t}{\tau} + \frac{\xi}{2}\right)\et^*\left(\frac{t}{\tau} - \frac{\xi}{2}\right)e^{i\omega \tau \xi}
\end{equation}

\begin{figure*}[ht]
\centering
\includegraphics[width = 2\columnwidth]{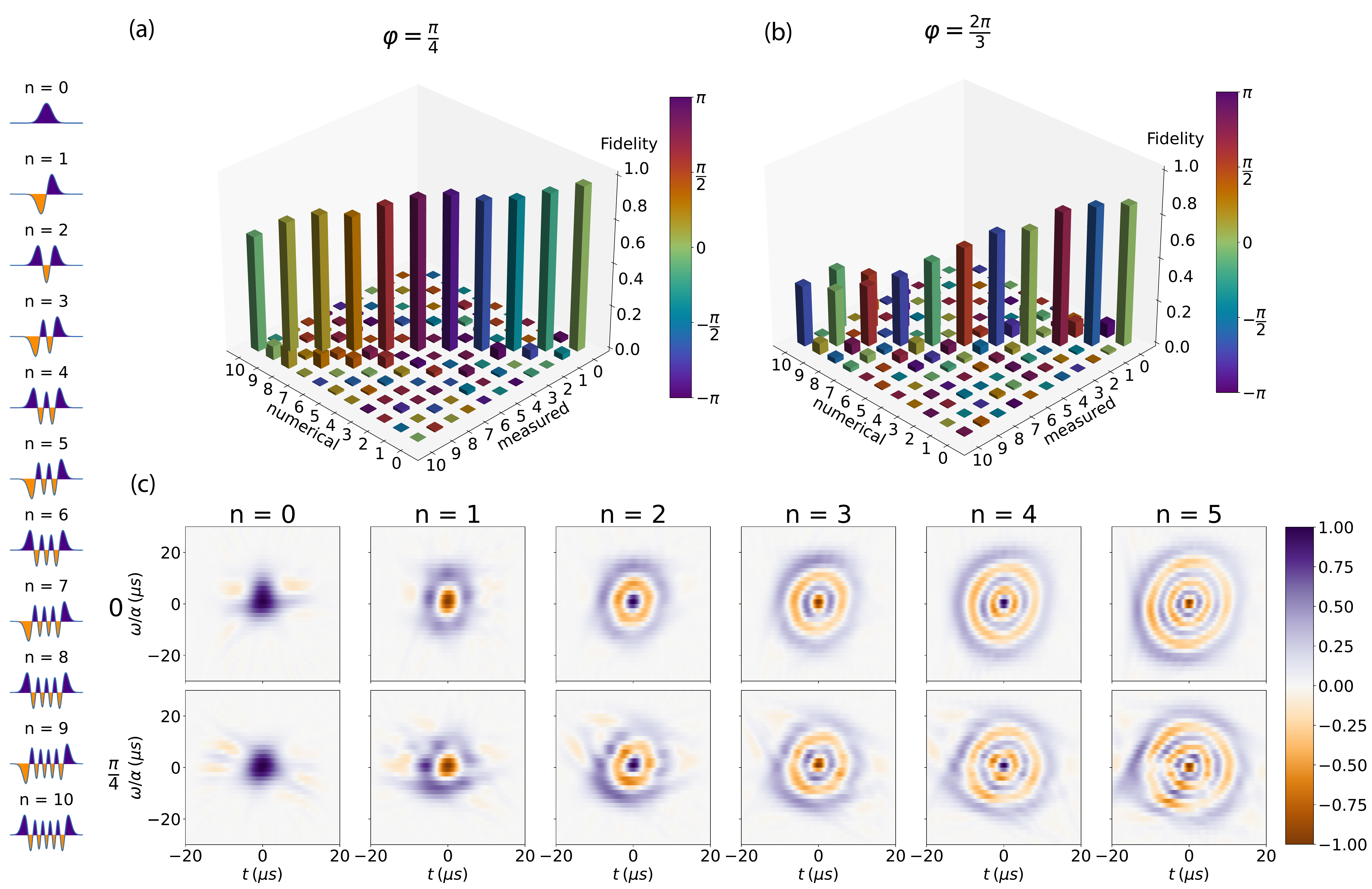}
\caption{\label{fig:gauss_hermite}
\subfig{a} Histogram of the fidelity for measured Hermite-Gaussian modes (measured) projected onto modes (numerical) for $\varphi =\frac{\pi}{4}$. The colormap represents the phase of the overlap.
\subfig{b} Histogram of the fidelity for measured Hermite-Gaussian modes (measured) projected onto modes (numerical) for $\varphi =\frac{2\pi}{3}$. The colormap represents the phase of the overlap.
\subfig{c} CWFs of measured subsequent Hermite-Gaussian modes for $n\in [0,10] \cap \mathbb{Z}$. The upper row represents $\varphi = 0$ and lower $\varphi = \frac{\pi}{4}$.}
\end{figure*}

The fractional Fourier transform applied to a pulse rotates its CWF by the angle $\varphi$. Such rotation can be achieved by adding quadratic time and frequency phases to the pulse \cite{Lohmann:93} in a sequence that represents two temporal lenses interleaved by a frequency lens with respective focusing powers $d_t$ and $d_\omega$. Such sequence of transformations applied to two Gaussian pulses is presented in the figure \ref{fig:idea}\subfig{a}. The lenses act on an optical signal as described below:
\begin{equation}\label{eq:lenses}
    \begin{split}
        \et(t/\tau)\xrightarrow{\text{temporal lens}}\et(t/\tau)\exp\left[-\frac{i d_{t}}{2}\left(\frac{t}{\tau}\right)^{2} \right], \\
        \Tilde{\et}(\omega \tau)\xrightarrow{\text{spectral lens}}\Tilde{\et}(\omega \tau)\exp\left[-\frac{id_\omega }{2}(\omega\tau)^2\right],
    \end{split}
\end{equation}
where $\Tilde{\et}(\omega\tau)$ is a Fourier transform of the optical signal $\et(t/\tau)$. The relation between $d_t$ and $d_\omega$ for an FrFT can be found by considering the general rotation matrix in the time-frequency domain \cite{Lohmann:93} and is given as follows:
\begin{equation}\label{eq:coeffs}
d_t(\varphi) = \tan\nicefrac{\varphi}{2} \hspace{15pt} d_\omega(\varphi)=\sin\varphi
\end{equation}

where $\varphi$ is the rotation angle. It is important to note that these equations hold only for angles $\varphi \in (-\pi, \pi)$. 
Adding each of the previously described phases to the pulse acts as shearing of the CWF of the signal as illustrated in Fig.~\ref{fig:idea}(a). 
Let $T$ and $B$ be the temporal and spectral widths. 
Action of the temporal lens leaves the temporal direction intact but broadens the bandwidth.
Therefore, the time-bandwidth area to be stored in the memory is given as:
\begin{equation}\label{bandwicz}
    TB' = TB(1+|\tan\nicefrac{\varphi}{2}|)
\end{equation} 
Eigenfunctions of FrFT are the Hermite-Gaussian functions $H_n^G(t/\tau)$ with eigenvalues being phasors with phase proportional to the angle of transformation $\varphi$ and order $n$, as described below.
\begin{equation}
\frft\left[H^{G}_{n}(t/\tau)\right] = H^{G}_{n}(t/\tau)e^{-in\varphi}
\label{eq:kermitFrFT}
\end{equation}

\paragraph{Experimental system}

\begin{figure*}[ht]
\centering
\includegraphics[width = 2\columnwidth]{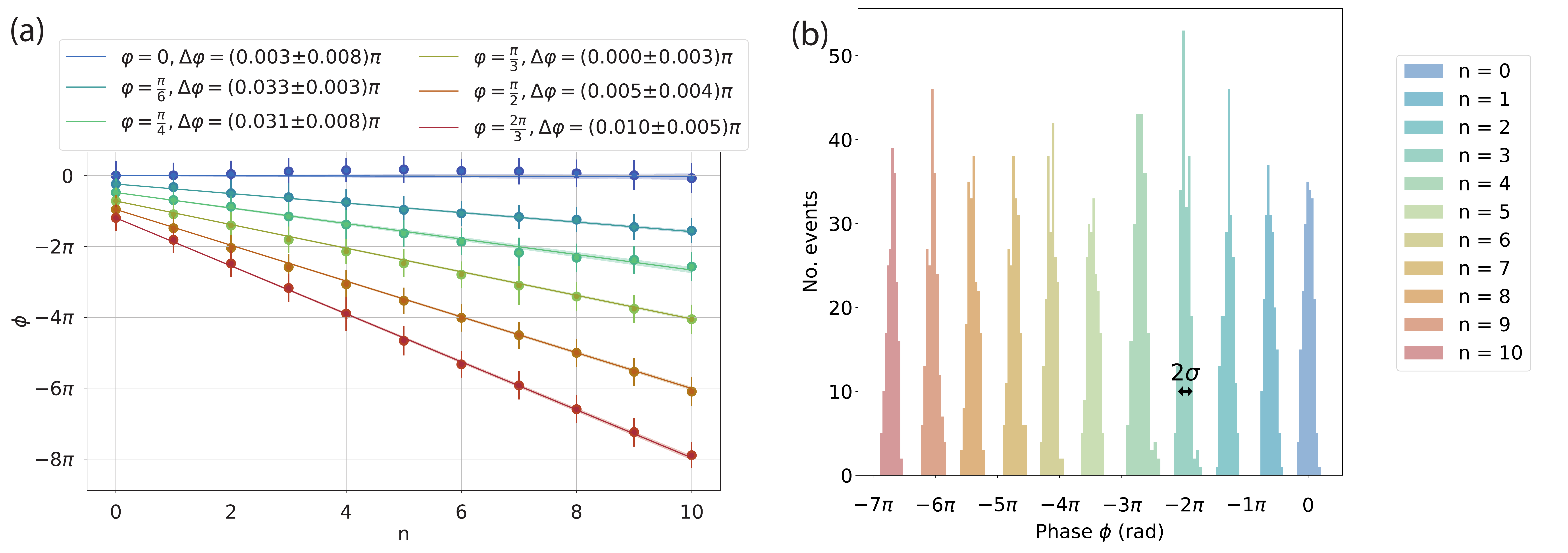}
\caption{\label{fig:wwlasne}\subfig{a} The phase of the overlap of subsequent Hermite-Gaussian modes for $n\in [0,10] \cap \mathbb{Z}$ with fitted functions. Error bars correspond to $5\sigma$. Shaded regions represent the fitted functions with $\pm\sigma$ of the fitted parameters. Each line is shifted by $\SI{0,75}{\rad}$ to clarify the plot.
\subfig{b} Histogram presenting phases of overlaps of Hermite-Gaussian modes $n\in [0,10] \cap \mathbb{Z}$ and $\varphi = \frac{2\pi}{3}$ with their numerical equivalents (the same mode, center, and width). The height of each bar is proportional to number of events i.e. measured phase of overlap is in a bin range. The bars are color-coded by each mode.}
\end{figure*}
The experiment is based on gradient echo quantum memory (GEM) that is built on rubidium-87 atoms trapped in a magneto-optical trap (MOT). Atoms form a cigarette shape \SI{10}{\mm} long cloud with an optical density reaching 85. The ensemble temperature is \SI{42}{\micro\K}. The setup schematic is presented in figure \ref{fig:idea}\subfig{c}. We utilize GEM protocol to map different frequencies of the signal to specific parts of the atomic cloud according to the formula $\omega = \beta\times z + \omega_{0}$, where $\beta$ is the value of the magnetic gradient, $z$ is the position along the $z$-axis and $\omega_{0}$ is the carrier frequency. We exploit that feature to add an arbitrary phase to the stored signal by the introduction of spatially varying ac-Stark shifts between the storage levels during the storage time. The spatial phase imposed by the beam onto the atomic state is equivalent to the spectral phase of a signal stored in the atomic cloud. To implement the frequency lens the ac-Stark beam is shaped to have a Fresnel-lens intensity profile imposing phase according to \eqref{eq:lenses}. The temporal lenses are imposed during the write-in and read-out process by linearly detuning the coupling laser in time at a matching rate of $d_t/\tau^2$. During the read-out we change the sign of detuning rate due to inverse time caused by different sign of magnetic gradient.
The main experimental limitation is the storage efficiency $\eta$ of the GEM protocol. 
Intense coupling laser is necessary for storage but at the same causes decoherence, thus effectively limiting the efficiency to $\eta = 1-\exp\left(-2\pi \mathrm{OD} / (TB')\right)$ \cite{Sparkes2013} where OD is the optical density of the atomic cloud. 
Prior to storage, the FrFT protocols broadens the bandwidth of the signal due to action of temporal lens as described in Eq. \eqref{bandwicz}.
In the experiment we picked $TB = 110$ to store first 11 Hermite-Gaussian modes. It follows that $T=\sqrt{110}\tau$ and $B=\sqrt{110}/\tau$. In our setup the time window is limited by power of coupling beam that induced a coherence decay at a rate of $\Gamma$ = \SI{9.1}{\kHz}, we chose $T\Gamma = 0.4$ obtaining $\eta = 33\%$, which sets the $\tau = \SI{4.2}{\us}$. The magnetic field gradient $\beta$ is adjusted to match desired storage bandwidth $\beta L=B'$. 
The read-out signal is sent to the homodyne detector connected to RedPitaya STEMlab 125-14 acting as an oscilloscope. For every measurement, we collect 200 homodyne waveforms of the read-out signal and average them by subtracting the phase of the LO from each measurement. The phase of the LO was inferred from additional impulse sent through after the read-out sequence. 
Moreover, we found that the GEM protocol itself imposes a constant quadratic frequency phase, which might be caused by eddy currents induced in passive elements surrounding the experimental setup. We have mitigated this issue by applying a constant frequency lens compensating for this unwanted effect. 
\paragraph{Results}

We start the demonstration of the FrFT by preparing a cat-like state input signal with an envelope:
\begin{equation}
    \psi^{\text{in}}(t/\tau) = \frac{\exp\left[\frac{-(t - \mu)^{2}}{2 (s\tau)^2}\right] + \exp\left[\frac{-(t + \mu)^{2}}{2 (s\tau)^{2}}\right]}{
    \sqrt{2 \sqrt{\pi} s\tau \left\{1 + \exp\left[-\mu^2 /(s\tau)^2\right]\right\}}}
\end{equation}
with $s\tau = \SI{2.4}{\us}$ and $\mu = \SI{7}{\us}$. Those parameters were chosen to efficiently use the whole available bandwidth and temporal window of the memory. The signal is then sent to the memory performing FrFT with chosen angles $\varphi\in\{0,\frac{\pi}{3},\frac{\pi}{2},\frac{2\pi}{3}\}$ The measured CWFs of the corresponding readout signals are depicted in Fig. \ref{fig:rotated_data} and compared with matching theoretical predictions. Achieved fidelity between simulation and experimental data varies between $65\%-88\%$.
Next, we benchmark the FrFT implementation by transforming Hermite-Gaussian input pulses with an envelope:
\begin{equation}
    \psi^{\text{in}}_n(t/\tau) = \mathrm{H}^G_n(t/\tau) = \frac{1}{\sqrt{2^{n}n!\tau\sqrt {\pi }}}H_{n}\left(\frac{t}{\tau}\right)\exp\left(\frac{-t^{2}}{2 \tau^{2}}\right)
\end{equation}
The quality of the FrFT is then probed by decomposing the measured readout signal for different FrFT angles $\varphi$ in the ideal (input) $\mathrm{H}^G_n$ basis with $n\in [0,10] \cap \mathbb{Z}$. 
Obtained complex decomposition coefficients arranged into transition matrix and presented in figures \ref{fig:gauss_hermite}\subfig{a} and \ref{fig:gauss_hermite}\subfig{b}. The coefficients are inferred from complex amplitude retrieved from homodyne measurement by calculating the overlap:
\begin{equation}
    \mathcal{F}_{n,m} = \frac{1}{\tau}\int dt \left[\mathrm{H}^G_n(t/\tau)\right]^*\psi^{\text{out}}_m(t/\tau)
\end{equation}
where $\psi^{\text{out}}_m$ is the signal measured at the output of the memory for set $m$, which is the index of the input Hermite-Gaussian mode and $n$ is the index of the projection mode. The height of the bars corresponds to $\left|\mathcal{F}_{n,m}\right|^2$, which is equivalent to the fidelity of measured signal and theoretical predictions. The color of the bars corresponds to the phase of the calculated overlap $\phi=\mathrm{arg}\left({\mathcal{F}_{n,m}}\right)$.
CWFs calculated from the measurements of the first 6 Hermite-Gaussian modes are shown in figure \ref{fig:gauss_hermite}\subfig{c}. After performing FrFT each Hermite-Gaussian mode gains phase $\phi$  proportional to the mode index $n$ as given by equation \eqref{eq:kermitFrFT}. These phases are plotted in figure \ref{fig:wwlasne}\subfig{a}. We fitted functions $\phi = \phi_{0} + n\frac{d\phi}{dn}$ to each set of the phases $\phi(n)$ of complex eigenvalues and obtained an effective FrFT angle. Our experiment yielded results presented in table \ref{tbl:results}. For each transformation, the set angle is denoted as $\varphi$, the measured rotation angle is $\nicefrac{d\phi}{dn}$, while their difference is represented as $\Delta\varphi$.
\begin{table}[ht]
\begin{tabularx}{\columnwidth} { 
  >{\centering\arraybackslash}X 
  | >{\centering\arraybackslash}X 
  | >{\centering\arraybackslash}X}
set angle $\varphi$&measured angle $\nicefrac{d\phi}{dn}$&deviation $\Delta\varphi$ [$\times 10^2$]\\
\hline
$0$&$0.001\pi$&$0.1 \pi$\\
$\pi/6$&$0.134\pi$&$-3.3 \pi$\\
$\pi/4$&$0.218\pi$&$-3.2 \pi$\\
$\pi/3$&$0.333\pi$&$-0.0 \pi$\\
$\pi/2$&$0.505\pi$&$0.5 \pi$\\
$2\pi/3$&$0.677\pi$&$1.0 \pi$\\
\end{tabularx}
\caption{Measured angles of FrFT for 11 consecutive Hermit-Gaussian mode impulses with their deviations from expected values and fit errors.}
\label{tbl:results}
\end{table}

In all cases, measured angles differ from desired by less than $0.033 \pi$. The overlap phase of every single measurement with generated Hermite-Gaussian mode for angle $\varphi=\frac{2\pi}{3}$ is shown on the histogram in figure \ref{fig:wwlasne}\subfig{b}. Obtained phase fluctuations are in order of $\SI{0.2}{\radian}$, which corresponds to a change in the value of the magnetic field by about $0.2\%$. 

\paragraph{Discussion}
In this letter, we tackled the problem of the physical implementation of the fractional Fourier Transform in the spectro-temporal domain. We leveraged the space-time duality and implemented the FrFT using the quantum-memory time-frequency processor. Utilizing a CWF we demonstrated a high-fidelity transformation of two-pulse states. Furthermore, we have characterized the implementation with the help of Hermite-Gaussian pulses -- eigenfunctions of FrFT -- and compared them with theoretical predictions. The results of this letter introduce many prospects for future research and applications, especially a method of sorting Hermite-Gaussian temporal modes utilizing FrFT \cite{Zhou2017}. 
Such a mode sorting technique is key for spectral superresolution with multi parameter quantum estimation \cite{Tsang2016}. The protocol presented in this letter opens exciting possibilities for implementation in the antipodal regime of ultrafast optics, where GEM protocol is not applicable and phases can be applied directly by dispersion fibers and EOMs. Despite of straightforward implementation, using ultrafast optics gives rise to new technical challenges not present in our version of the protocol, but likely to spark interest of the ultrafast optics community.
\section{Acknowledgments}
\begin{acknowledgments}
The “Quantum Optical Technologies” (MAB/2018/4) project is carried out within the International Research Agendas programme of the Foundation for Polish Science co-financed by the European Union under the European Regional Development Fund. We thank K. Banaszek for generous support.
\end{acknowledgments}
\bibliographystyle{apsrev4-2}
\bibliography{refs}
\end{document}